\documentclass[preprint, english]{revtex4-1}
\usepackage[T1]{fontenc}
\usepackage[utf8]{luainputenc}
\usepackage{amsmath,amsfonts,amsthm,bm}
\usepackage{amssymb}
\usepackage{graphicx}
\usepackage{esint}
\usepackage{babel}

\begin{document}

\title{Modular Operators and Entanglement in Supersymmetric Quantum Mechanics}

\author{Rupak Chatterjee}
\email{corresponding author: Rupak.Chatterjee@Stevens.edu}
\author{Ting Yu}

\affiliation{Center for Quantum Science and Engineering\\
Department of Physics, Stevens Institute of Technology, Castle Point on the Hudson, Hoboken, NJ 07030}

\begin{abstract}
The modular operator approach of Tomita-Takesaki to von Neumann algebras is elucidated in the algebraic structure of certain supersymmetric quantum mechanical systems. A von Neumann algebra is constructed from the operators of the system.  An explicit  operator characterizing the dual infinite degeneracy structure of a supersymmetric two dimensional system is given by the modular conjugation operator. Furthermore, the entanglement of formation for these supersymmetric systems using concurrence is shown to be related to the expectation value of the modular conjugation operator in an entangled bi-partite supermultiplet state thus providing a direct physical meaning to this anti-unitary, anti-linear operator as a quantitative measure of entanglement.  Finally, the theory is applied to the case of two-dimensional Dirac fermions, as is found in graphene, and a  supersymmetric Jaynes Cummings Model.  

\end{abstract}

\maketitle

\section{Introduction}
Supersymmetric (SUSY) quantum mechanics \cite{Junk1996} has shown to be a useful arena in which to invesitigate  various mathematical and physical properties of symmetry operators throughout diverse fields of theoretical physics \cite{Oikonomou2011, Oikonomou2013,Oikonomou2014, And2003, And2012}. Here, we extend the available symmetry operators to include the modular operators first introduced by Tomita and Takesaki \cite{Tak1970}. The  modular operator approach to von Neumann $C^*$-algebras has already found various uses in the field of algebraic quantum field theory \cite{Zeidler2009, Haag1996, Baumgartel1992}. In this paper, we give a physical meaning of the modular conjugation operator as a quantitative measure of entanglement for bi-partite states. As described in \cite{Uhlmann2000}, an anti-linear, anti-unitary operator is the key driving force behind the entanglement of formation for bipartite entangled systems.

Two dimensional Dirac fermions and the closely related phenomena of Landau levels created by the motion of planar  electrons in a constant magnetic field has been successfully applied in understanding certain properties of graphene. One previous study \cite{Ali2010} investigated the modular operator approach of von Neumann $C^*$-algebras to non-supersymmetric Landau levels. Here, we extend this analysis in section II to a supersymmetric case. We follow a supersymmetric structure of quantum Hall effects in graphene investigated in \cite{Ezawa2008}. Other works in analyzing supersymmetric structures in graphene may be found in \cite{Midya2014, Halberg2017, Abreu2011}. Here, rather than investigating a new supersymmetric structure, we further elucidate the algebraic structure of supersymmetric Landau levels from the perspective of a von Neumann algebra of operators and their corresponding modular operator structure. The Tomita Takesaki modular conjugation operator characterizing the dual infinite degeneracy of the supersymmetric Hamiltonian is the key result of section III. We first elucidate the structure in a holomorphic representation and thereafter provide the general definition of the modular conjugation operator.

In section IV, the entanglement of formation is investigated for an energy supermultiplet of a SUSY quantum system using concurrence. The main result is that the concurrence of an entangled bi-partite supermultiplet state is related to the expectation value of the Tomita Takesaki modular conjugation operator in that state thus providing a direct physical meaning to this anti-unitary, anti-linear operator as a quantitative measure of entanglement. The algebraic structure of section III is then applied in section V to a Hamiltonian describing the massless low energy excitations near the two Dirac points of the Brillouin zone of graphene. Utilizing a modular conjugation operator, we relate various different Hamiltonians at Dirac points in the first Brillouin zone. Finally, in section VI, we investigate the modular properties of the SUSY algebraic structure of the Jaynes Cummings (JC) model in resonance. The JC model in the rotating wave approximation (RWA) is mapped by a modular conjugation operator into an anti-JC model with non-RWA interaction terms.  Finally, the appendix lists various mathematical results in the theory of von Neumann algebras and their modular operator structure along with definitions of SUSY quantum mechanics and the entanglement of formation and concurrence. It also includes some detailed proofs left out of the main body of the paper.

\section{Landau Levels and Supersymmetry}

Consider the well known Hamiltonian \cite{Landaul1977} of an electron in a magnetic field
\begin{equation}
H = \dfrac{1}{2m} \mathbf{\Pi} ^2 -\boldsymbol{\mu} \cdot\mathbf{B},
\end{equation}
where we have used the Peierls substitution $ \mathbf{\Pi} = \mathbf{p} - \dfrac{e}{c} \mathbf{A}$.
Here, we restrict the motion to a two-dimensional $x-y$ plane and assume the magnetic field is perpendicular to this plane, $ \mathbf{B} = (0,0, B)$. The magnetic moment is given by $
\boldsymbol{\mu} = g \dfrac{e\mathbf{s}}{2mc}=g \dfrac{e\hbar \boldsymbol{\sigma}}{4mc}$
where we have used the Pauli matrix prescription $\mathbf{s} = \hbar \boldsymbol{\sigma}/2$.
The standard quantization prescription is initiated by introducing creation and annihilation operators related to the quantized Peierls momenta,  
\begin{equation}
[\Pi_x , \Pi_y] = i \dfrac{e\hbar}{c} B ,\;\;\;
\\
a = \sqrt{\dfrac{c}{2e\hbar B}} (i\Pi_x - \Pi_y ) ,\;\;\;
\\
a^\dagger = \sqrt{\dfrac{c}{2e\hbar B}} (-i\Pi_x - \Pi_y ) .
\end{equation}
Consider an electron with $g \approx 2$. Using $[a,a^\dagger]=1$ and defining the frequency $\omega = eB/mc$, (1) becomes
\begin{equation}
H^a = \hbar \omega \left( a^\dagger a  + \dfrac{1}{2} -\dfrac{\hat{\sigma}_z}{2}  \right).
\end{equation}
The Hilbert space for this Hamiltonian  will be denoted by $\mathcal{H} = \mathcal{F} \otimes \mathbb{C}^2$ where $ |n \rangle \in \mathcal{F}$ is the Bosonic Fock space and $\alpha 
\begin{pmatrix}
1 \\
0
\end{pmatrix} 
+\gamma \begin{pmatrix}
0 \\
1
\end{pmatrix} 
 \in \mathbb{C}^2$. The Landau energy levels easily follow as
\begin{equation}
E^a_n = \hbar \omega \left( n  + \dfrac{1}{2} - \dfrac{\sigma_z}{2} \right), \,\,\, \sigma_z = \pm 1, \,\,\, n \in \mathbb{Z}^{0+}.
\end{equation}
The ground state energy $n=0$ with "spin up" $\sigma_z =1$,  is precisely zero $E^a_0 (\sigma_z =1)=0$, whereas the excited energy levels  $E^a_n$  are two-fold degenerate 
\begin{equation}
E^a_n =  \hbar \omega k \,\,\, \Bigg\{
\begin{array}{c}
 n=k-1, \sigma_z = -1\\
 n=k, \sigma_z =  1\\
\end{array} , \,\,\, k \in \mathbb{Z}^{+}.
\end{equation}
This degeneracy along with the unique ground state energy can be described by a supercharge operator $Q$ of SUSY quantum mechanics \cite {Witt1981}. $Q$ annihilates the ground state ($E_0=0$), and is related to the Hamiltonian by the anti-commutation relation $H = \hbar \omega \{Q, Q^\dagger \} $ as defined in the appendix. There, $Q$ is shown to commute with the supersymmetric Hamiltonian $H$ thereby getting the name "supercharge" analogous to conserved charges in quantum field theory.   

For our system, the supersymmetric charges are given by
\begin{equation}
Q^a = a  \otimes \sigma_- = 
\begin{pmatrix}
0 & 0 \\
a & 0
\end{pmatrix}, \;\;
\;\;
Q^{a \dagger} = a^\dagger  \otimes \sigma_+ =
\begin{pmatrix}
0 & a^\dagger \\
0 & 0
\end{pmatrix},
\end{equation}
where
\begin{equation}
H^a / \hbar \omega = \{ Q^a, Q^{a \dagger} \} 
= \left(a^\dagger a +\dfrac{1}{2} \right) \otimes I - \hat{1} \otimes \dfrac{1}{2} \hat{\sigma}_z ,
\end{equation}
as equation (3) requires. The superalgebra is easily verified to be $H^a = \hbar \omega \{Q^a, Q^{a \dagger} \}$ and $\{ Q^a, Q^a \}  = \{ Q^{a \dagger},  Q^{a \dagger} \} =[H^a, Q^a]=[H^a, Q^{a \dagger}] = 0$.
The conserved supercharge operator that creates the SUSY energy degeneracy, $Q^a_{SUSY} = Q^a+ Q^{a \dagger}$ takes a spin-down state $\Phi^{k-1}_{\downarrow} \in \mathcal{H}$ to a higher excited Landau level and raises its spin, i.e.
\begin{equation}
Q^a_{SUSY}  \Phi^{k-1}_{\downarrow}   =
\begin{pmatrix}
0 & a^\dagger \\
a & 0
\end{pmatrix}
\begin{pmatrix}
0  \\
|k-1 \rangle
\end{pmatrix} = \sqrt{k}
\begin{pmatrix}
|k\rangle  \\
0
\end{pmatrix} =\sqrt{k} \Phi^{k}_{\uparrow} ,
\end{equation}
or takes a spin-up state (other than the ground state) to a lower Landau level and lowers its spin, 
\begin{equation}
Q^a_{SUSY}  \Phi^{k}_{\uparrow}   =
\begin{pmatrix}
0 & a^\dagger \\
a & 0
\end{pmatrix}
\begin{pmatrix}
|k \rangle \\
0
\end{pmatrix} = \sqrt{k}
\begin{pmatrix}
0  \\
|k-1 \rangle
\end{pmatrix} 
=\sqrt{k}  \Phi^{k-1}_{\downarrow},
\end{equation}
where the energy degeneracy can be read from equation (5) since it is easy to see that $H^a = \hbar \omega (Q^a_{SUSY})^2$ . Therefore, the two states
\begin{equation}
\left\{ \Phi^{k}_{\uparrow} ,\Phi^{k-1}_{\downarrow} \right\} =
\left\{
\begin{pmatrix}
|k \rangle \\
0
\end{pmatrix} ,
\begin{pmatrix}
0  \\
|k-1 \rangle
\end{pmatrix}
\right\}
\end{equation}
form the supermultiplet with energy $E^a_k  = \hbar \omega k$.

The classical mechanical description of an electron in an external magnetic field leads to cyclotron motion of the charged particle. Furthermore, the classical mechanical energy does not depend on the position of the center of cyclotron motion thereby leading to a constant of motion. This center remains a constant of motion in the quantum mechanical case also. The center of cyclotron motion has quantized coordinates given by  \cite{Ezawa2013}
\begin{equation}
X= x +\dfrac{c}{eB} \left( p_y - \dfrac{e}{c} A_y \right) =  x + \dfrac{\Pi_y}{mw} ,\;\;
Y= y -\dfrac{c}{eB} \left( p_x - \dfrac{e}{c} A_x \right) =  y - \dfrac{\Pi_x}{mw} ,
\end{equation}
with a commutation relation given by $[X,Y] = -\dfrac{i \hbar c}{eB}$.
The guiding center $(X,Y)$ and the Peierls momenta $(\Pi_x , \Pi_y )$ form two commuting sets of operators, i.e.
\begin{equation}
\, [X,\Pi_x] = [X,\Pi_y] = 0,\;\;
\, [Y,\Pi_x] = [Y,\Pi_y] = 0.
\end{equation}
By introducing another set of bosonic creation -annihilation operators, 
\begin{equation}
b = \sqrt{\dfrac{eB}{2\hbar c}} (X - iY), \;\;
\\
b^\dagger = \sqrt{\dfrac{eB}{2\hbar c}} (X + iY), \;\;
\,[b, b^\dagger]=1,
\end{equation}
one may find the well known degeneracy of each Landau level (related to the angular momentum of cyclotron motion) and the degeneracy "filling factor" for finite systems (see \cite{Jain2007}). As the '$b$' operators commute with '$a$' operators, the supersymmetric construction above is retained. This degeneracy will be further elucidated in the following section.

\section{Von Neumann Algebras and Modular Operators}
Here, we investigate the algebraic structure of the supersymmtric  Landau levels above from the perspective of von Neumann algebras and their corresponding modular operator structure.
Consider the symmetric gauge $\mathbf{A} = \dfrac{B}{2} (-y, x, 0), \,\,\ \mathbf{B}= \nabla \times \mathbf{A} = (0,0,B)$ whereby the  operators (13) can be re-written as 
\begin{equation}
b = \sqrt{\dfrac{c}{2e\hbar B}} (i \tilde{\Pi}_x  - \tilde{\Pi}_y ), 
\;\;
b^\dagger = \sqrt{\dfrac{c}{2e\hbar B}} (-i \tilde{\Pi}_x  - \tilde{\Pi}_y),
\end{equation}
where $ \mathbf{\tilde{\Pi}} = \mathbf{p} + \dfrac{e}{c} \mathbf{A}$ and $ [\tilde{\Pi}_x , \tilde{\Pi}_y] = -i \dfrac{e\hbar}{c} B$ (note that the sign of the second term has changed with respect to the Peierls momenta $\mathbf{\Pi}$). As operators (14) commute with the Hamiltonian (7) the explicit degeneracy of the now composite bosonic Fock space $\mathcal{F}$ of $\mathcal{H} = \mathcal{F} \otimes \mathbb{C}^2$, $|n,m \rangle \in \mathcal{F}$ can be written as  
\begin{equation}
|n,m \rangle = \dfrac{{a^\dagger}^n {b^\dagger}^m}{\sqrt{n!m!}} |0,0 \rangle , n,m \in \mathbb{Z}^{0+} .
\end{equation}
The energy of these states is still given by (5) which being independent of $m$ indicates an infinite degeneracy per $n^{th}$ Landau level. 

Due to the similar form of (14) compared to (2), we can reinterpret $ \mathbf{\tilde{\Pi}} = \mathbf{p} + \dfrac{e}{c} \mathbf{A}$ as a Peierls operator $ \mathbf{\tilde{\Pi}} = \mathbf{p} - \dfrac{e}{c} \mathbf{\tilde{A}}$ where $\mathbf{\tilde{A}} = -\mathbf{A} = \dfrac{B}{2} (y, -x, 0), \,\,\ \mathbf{\tilde{B}}= \nabla \times \mathbf{\tilde{A}} = (0,0,-B)$ is a symmetric gauge with the magnetic field pointing in the negative $z$-direction. In this way, the '$b$' operators become fundamental with the '$a$' operators creating the infinite degeneracy of the energy levels being given by (5) with $n$ replaced by $m$. We will denote these two cases of positive and negative $z$-direction $B$-field systems by the superscripts '$a$' and '$b$'. The algebraic structure of the  '$b$' system is similar to the '$a$' system of section II, 
\begin{equation}
Q^{b \dagger} = b^\dagger  \otimes \sigma_- =
\begin{pmatrix}
0 & 0 \\
b^\dagger & 0
\end{pmatrix}, \;\;
\\
Q^{b} = b  \otimes \sigma_+ = 
\begin{pmatrix}
0 & b \\
0 & 0
\end{pmatrix},
\end{equation}
where
\begin{equation}
H^b / \hbar \omega = \{ Q^b , Q^{b \dagger} \} 
= \left(b^\dagger b +\dfrac{1}{2} \right)\otimes I + \hat{1} \otimes \dfrac{1}{2} \hat{\sigma}_z .
\end{equation}
If the magnetic field points in the negative $z$-direction, this is the Hamiltonian with a (supersymmetric) degeneracy of  
\begin{equation}
\begin{array}{c}
E^b_m = \hbar \omega \left( m + \dfrac{1}{2} + \dfrac{\sigma_z}{2} \right), \,\,\, \sigma_z = \pm 1,\,\,\, m \in \mathbb{Z}^{0+} , \\
E^b_m =  \hbar \omega m \,\,\, \Bigg\{
\begin{array}{c}
m=l-1, \sigma_z = 1\\
m=l, \sigma_z =  -1\\
\end{array} , \,\,\, l \in \mathbb{Z}^{+} .\\
\end{array}
\end{equation}
Here, the ground state energy $m=0$ with "spin down" $\sigma_z =-1$,  is precisely zero $E^b_0(\sigma_z = -1)=0$.

The conserved supercharge operator that creates the SUSY energy degeneracy, $Q^b_{SUSY} = Q^b+ Q^{b\dagger}$ takes a spin-down state (other than the ground state) $\Phi^{l}_{\downarrow} \in \mathcal{H}$ to a lower excited Landau level and raises its spin, i.e.
\begin{equation}
Q^b_{SUSY} \Phi^{l}_{\downarrow}   =
\begin{pmatrix}
0 & b \\
b^\dagger & 0
\end{pmatrix}
\begin{pmatrix}
0  \\
|l \rangle
\end{pmatrix} = \sqrt{l}
\begin{pmatrix}
|l-1\rangle  \\
0
\end{pmatrix} =\sqrt{l} \Phi^{l-1}_{\uparrow} 
\end{equation}
or takes a spin-up state to a higher Landau level and lowers its spin, 
\begin{equation}
Q^b_{SUSY} \Phi^{l-1}_{\uparrow}   =
\begin{pmatrix}
0 & b \\
b^\dagger & 0
\end{pmatrix}
\begin{pmatrix}
|l-1 \rangle \\
0
\end{pmatrix} = \sqrt{l}
\begin{pmatrix}
0  \\
|l \rangle
\end{pmatrix} 
=\sqrt{l}  \Phi^{l}_{\downarrow}
\end{equation}
where the energy degeneracy can be read from equation (18) since $H^b = \hbar \omega 
( Q^b_{SUSY} )^2$. Therefore, the two states 
\begin{equation}
\left\{ \Phi^{l-1}_{\uparrow} ,\Phi^{l}_{\downarrow} \right\} =
\left\{
\begin{pmatrix}
|l-1 \rangle \\
0
\end{pmatrix} ,
\begin{pmatrix}
0  \\
|l \rangle
\end{pmatrix}
\right\}
\end{equation}
form the supermultiplet with energy $E^b_l =  \hbar \omega l $. The  supersymmetric algebra combining both the above systems is
\begin{equation}
H^a = \hbar \omega \{Q^a, Q^{a \dagger} \}, \;\;
H^b = \hbar \omega \{Q^b, Q^{b \dagger} \} ,\;\;
[H^a, H^b]=0,
\end{equation}
where the remaining commutation and anti-commutation relations, i.e. $\{ Q^a, Q^a \}, \{ Q^{a \dagger}, Q^{b \dagger} \}, ...$ are all zero.

The symmetry between these two related systems of commuting algebras may be generated by Tomita-Takesaki modular operators from a standard form of von Neumann algebras (see the appendix and  \cite{Haag1996}). Let $\mathcal{A}$ be a representation of a von Neumann algebra of bounded linear operators  on a Hilbert space $\mathcal{H}$, $\mathcal{A} \subset \mathcal{B(H)}$.  $\mathcal{A}$ is a $C^*$-algebra with a commutant $\mathcal{A}'$ (the set of elements in $\mathcal{B(H)}$ commuting with $\mathcal{A}$) such that $\mathcal{A}''=\mathcal{A}$ (the double commutant returns one to $\mathcal{A}$ as defined in the appendix). Let $|\Omega \rangle  \in \mathcal{H} $ be a separating and cyclic vector for the $C^*$-algebra $\mathcal{A}$ (the elements in  $\mathcal{A}$ can create a space dense in $\mathcal{H}$ by acting on $|\Omega \rangle  \in \mathcal{H} $). There exists an anti-linear map $ S: \mathcal{H} \rightarrow \mathcal{H},$ such that $ S A |\Omega \rangle  = A^* |\Omega \rangle, \forall A \in \mathcal{A} $. $S$ has a polar decomposition given by $S = J \Delta^{1/2} =  \Delta^{-1/2} J, \,\, \Delta = S^{*} S$ where the modular conjugation operator $J$ has the following properties (see appendix),
\begin{equation}
J \Delta ^{\frac{1}{2}} J = \Delta ^{-\frac{1}{2}}, \;\;\;
J^2 =I, \,\,\, J^{*} = J, \;\;\;
J |\Omega \rangle  = |\Omega \rangle,  \;\;\;
J \mathcal{A} J = \mathcal{A}' .
\end{equation}
The key feature for our purposes is that the modular conjugation operator $J$ takes an algebra $\mathcal{A}$ into its commutant $\mathcal{A}'$. The von Neumann algebras considered here will have elements of the form of unitary Weyl operators of the supercharge generators $Q^a, Q^{a \dagger}, Q^b, Q^{b \dagger}$,
\begin{equation}
\Big\{ \exp[-i(\alpha Q^a +\beta Q^{a \dagger} )] \Big\} \subset \mathcal{A}, \;\;
\Big\{ \exp[-i(\gamma Q^b +\delta Q^{b \dagger} )] \Big\} \subset \mathcal{A}' ,\;\;
\alpha, \beta,  \gamma, \delta \in \mathbb{R}.
\end{equation}
The $*$-operation of the $C^*$-algebra structure is given by the adjoint operation $\dagger$. Note that due to the algebraic structure of the super charges, a formal power series expansion of an element of $\mathcal{A}$ (or analogously $\mathcal{A}'$)  is given by $\exp[-i(\alpha Q^a +\beta Q^{a \dagger})] = 1 - i\alpha Q^a - i\beta Q^{a \dagger} - \frac{\alpha \beta}{2 \hbar \omega}H^a  $. Therefore, we restrict our analysis below to the supercharges and Hamiltonians of our two systems.

Introducing a magnetic length $l_m = \sqrt{\hbar c / {eB}}$, the matrix elements of the supercharge generators of our two algebras are given by 
\begin{equation}
a = \dfrac{l_m}{\sqrt{2} \hbar} (i\Pi_x - \Pi_y ) , \,\,\ a^\dagger = \dfrac{l_m}{\sqrt{2} \hbar} (-i\Pi_x - \Pi_y ) , \;\;
b = \dfrac{1}{\sqrt{2} l_m} (X - iY ) , \,\,\ b^\dagger = \dfrac{1}{\sqrt{2} l_m} (X +i Y) . 
\end{equation}
Using the Schr\"{o}dinger representation in the symmetric gauge
\begin{equation}
\Pi_x = -i\hbar\dfrac{\partial}{\partial x} +\dfrac{\hbar}{2 l_m^2} y , \,\,\ \Pi_y = -i\hbar\dfrac{\partial}{\partial y} - \dfrac{\hbar}{2 l_m^2} x , \;
\;
X = \dfrac{1}{2}x -il_m^2 \dfrac{\partial}{\partial y} , \,\,\ Y = \dfrac{1}{2}y +il_m^2 \dfrac{\partial}{\partial x},
\end{equation}
and complex co-ordinates for a holomorphic representation (see, for example, the recent review \cite{Roug2020})
\begin{equation}
z = \dfrac{1}{2l_m}(x+iy), \,\,\, \bar{z} = \dfrac{1}{2l_m}(x-iy), \;
\;
\dfrac{\partial}{\partial z} = l_m \left( \dfrac{\partial}{\partial x} -i\dfrac{\partial}{\partial y} \right) , \,\,\ \dfrac{\partial}{\partial \bar{z}} = l_m \left( \dfrac{\partial}{\partial x} + i\dfrac{\partial}{\partial y} \right),
\end{equation}
our operators become
\begin{equation}
a = \dfrac{1}{\sqrt{2}}\left(z+\dfrac{\partial}{\partial \bar{z}} \right), \,\,\, a^\dagger = \dfrac{1}{\sqrt{2}}\left(\bar{z} - \dfrac{\partial}{\partial z} \right), \;
\;
b = \dfrac{1}{\sqrt{2}}\left(\bar{z}+\dfrac{\partial}{\partial z} \right), \,\,\, b^\dagger = \dfrac{1}{\sqrt{2}}\left(z - \dfrac{\partial}{\partial \bar{z}} \right). 
\end{equation}

Note that the '$a$'s and '$b$'s are related by complex conjugation.
These operators work on the  Fock space component of the Hilbert Space $\mathcal{H} = \mathcal{F} \otimes \mathbb{C}^2$.  Wavefunctions in the holomorphic representation above will be of the form $ \psi_{nm}(z, \bar{z}) = \langle z, \bar{z} | n,m \rangle \in \mathcal{F} $. We define our modular conjugation operator $J$ to act as follows. Let $J: \mathcal{F} \otimes \mathbb{C}^2 \rightarrow \mathcal{F} \otimes \mathbb{C}^2 $. Then, the modular conjugation operator is given by
\begin{equation}
J \left [ \psi_{nm}(z, \bar{z}) \otimes \left\{ \alpha
\begin{pmatrix}
1\\
0
\end{pmatrix}  + \beta 
\begin{pmatrix}
0\\
1
\end{pmatrix}  \right\}  \right] 
= \left [ \bar{\psi}_{mn}(z, \bar{z}) \otimes \left\{ \bar{\alpha}
\begin{pmatrix}
0\\
1
\end{pmatrix}  + \bar{\beta} 
\begin{pmatrix}
1\\
0
\end{pmatrix}  \right\}  \right]  
\end{equation}
The modular conjugation operator $J$ acts as complex conjugation while exchanging the two indices of the composite bosonic Fock space and performing a spin flip in $\mathbb{C}^2$.  

We now prove some key elements of our Tomita-Takesaki modular operator structure leaving complete proofs to the general definition of $J$ given below. From the definition above of our von Neumann algebra elements, our $*$-operation is given by the adjoint operator $\dagger$. Consider the $*$-operation on one of the generators of $\mathcal{A}$, $Q^a$,
\begin{equation}
(Q^{a})^* =(Q^{a})^{\dagger}=
\begin{pmatrix}
0 & 0 \\
a & 0
\end{pmatrix}^{\dagger}=
\begin{pmatrix}
0 & \bar{a} \\
0 & 0
\end{pmatrix}= 
\begin{pmatrix}
0 & b \\
0 & 0
\end{pmatrix} = Q^b
\end{equation}
where we have used the relations in (28). Clearly, operating with $\dagger$ again returns us to the original generator. The other generators follow a similar pattern indicating that the generators of $\mathcal{A}$ are mapped into the commutant $\mathcal{A}'$ via the $*$-operation and $\mathcal{A}''= \mathcal{A} $ for our von Neumann algebra. The main aspect of our modular conjugation operator is that $J$ maps $\mathcal{A}$ into $\mathcal{A}'$, i.e. $J \mathcal{A} J = \mathcal{A}'$. Acting on $\mathcal{F}$ , one finds, for instance that
\begin{equation}
\left\{ J a J \right\} \psi_{nm}(z, \bar{z}) 
= J  \dfrac{1}{\sqrt{2}}\left(z+\dfrac{\partial}{\partial \bar{z}} \right) \bar{\psi}_{mn}(z, \bar{z}) 
= \dfrac{1}{\sqrt{2}} \left( \bar{z} \psi_{nm}(z, \bar{z})+\dfrac{\partial \psi_{nm}(z, \bar{z})}{\partial z} \right) 
 = b  \psi_{nm}(z, \bar{z}) .
\end{equation}
Furthermore,  
\begin{equation}
\begin{array}{c}
J Q^a J 
\begin{pmatrix}
0\\
\bar{\psi}_{nm}(z, \bar{z})
\end{pmatrix} 
= J \begin{pmatrix}
0\\
a \psi_{mn}(z, \bar{z})
\end{pmatrix} 
= J  
\begin{pmatrix}
0\\
\dfrac{1}{\sqrt{2}}\left(z+\dfrac{\partial}{\partial \bar{z}} \right) \psi_{mn}(z,\bar{z})
\end{pmatrix} \\ \\  
= \begin{pmatrix}
b \bar{\psi}_{nm}(z, \bar{z}) \\
0 
\end{pmatrix} 
=
\begin{pmatrix}
0 & b\\
0 & 0
\end{pmatrix}
\begin{pmatrix}
0\\
\bar{\psi}_{nm}(z, \bar{z})
\end{pmatrix} = Q^{b}
\begin{pmatrix}
0\\
\bar{\psi}_{nm}(z, \bar{z})
\end{pmatrix} .
\end{array}
\end{equation}
A similar calculation holds for the other three operators, and therefore $J \mathcal{A} J = \mathcal{A}'$. Furthermore, using (17)
\begin{equation}
\begin{array}{c}
J [H^b /{\hbar \omega}] J \left [ \psi_{nm}(z, \bar{z}) \otimes 
\begin{pmatrix}
1\\
0
\end{pmatrix} \right] = 
J \left\{ \left( b^\dagger b + \dfrac{1}{2} \right) \otimes  I + \hat{1} \otimes \dfrac{\hat{\sigma}_z}{2} \right\} \left[ \bar{\psi}_{mn}(z, \bar{z}) \otimes 
\begin{pmatrix}
0 \\
1
\end{pmatrix} \right]  \\
= J \left\{ b^\dagger b \bar{\psi}_{mn}(z, \bar{z}) \otimes 
\begin{pmatrix}
0 \\
1
\end{pmatrix}  + \dfrac{\bar{\psi}_{mn}(z,\bar{z})}{2} \otimes 
\begin{pmatrix}
0 \\
1
\end{pmatrix} -
\dfrac{\bar{\psi}_{mn}(z, \bar{z})}{2} \otimes 
\begin{pmatrix}
0 \\
1
\end{pmatrix} 
\right\} \\
= J \left\{ b^\dagger b \bar{\psi}_{mn}(z, \bar{z}) \otimes 
\begin{pmatrix}
0 \\
1
\end{pmatrix} \right\} = a^\dagger a \psi_{nm}(z, \bar{z}) \otimes 
\begin{pmatrix}
1 \\
0
\end{pmatrix} \\
= \left\{ \left( a^\dagger a + \dfrac{1}{2} \right) \otimes  I - 1 \otimes \dfrac{\hat{\sigma}_z}{2} \right\} \left[ \psi_{nm}(z, \bar{z}) \otimes 
\begin{pmatrix}
1 \\
0
\end{pmatrix} \right] 
= [H^a /{\hbar \omega}] \left [ \psi_{nm}(z, \bar{z}) \otimes 
\begin{pmatrix}
1\\
0
\end{pmatrix} \right]

\end{array}
\end{equation}
and therefore, $J H^b J = H^a$. $J H^a J = H^b$ follows in a similar manner. $J$ maps the two supersymmetric systems, with reversed magnetic field directions, into one another thereby providing an explicit symmetry operator. With a fixed magnetic field, one may see $J$ as a map between the Landau level operators and quantized coordinates of the center of cyclotron motion. 

The above action of the modular conjugation operator $J$ was for the special case of a holomorphic representation in the symmetric gauge. In general, the modular conjugation operator is given as follows. Let $J: \mathcal{F} \otimes \mathbb{C}^2 \rightarrow \mathcal{F} \otimes \mathbb{C}^2 $. Then, the modular conjugation operator is given by
\begin{equation}
J \left[ |n , m \rangle \otimes \left\{ \alpha
\begin{pmatrix}
1 \\
0
\end{pmatrix} +   \beta  \begin{pmatrix}
0 \\
1
\end{pmatrix} \right\} \right] = 
\left[ |m, n \rangle \otimes \left\{ \bar{\alpha}
\begin{pmatrix}
0 \\
1
\end{pmatrix} +   \bar{\beta}  \begin{pmatrix}
1 \\
0
\end{pmatrix} \right\} \right]
\end{equation}
The basic properties for $J$ stated in (23) are verified in the appendix.

\section{Modular Operators and Concurrence for the Entanglement of Formation}
Given the Hilbert Space $\mathcal{H}^a = \mathcal{F}^a \otimes \mathbb{C}^2$, the Hamiltonian $H^a  = \hbar \omega \{ Q^a, Q^{a \dagger} \}$
has separable eigenstates of the supermultiplet 
\begin{equation}
\left\{ |k \rangle \otimes
\begin{pmatrix}
1 \\
0
\end{pmatrix} ,
|k-1 \rangle \otimes
\begin{pmatrix}
0  \\
1
\end{pmatrix}
\right\},
\end{equation}
or the (maximally) entangled state  
\begin{equation}
\dfrac{1}{\sqrt{2}}
\begin{pmatrix}
|k \rangle \\
\pm |k-1 \rangle
\end{pmatrix} = \dfrac{1}{\sqrt{2}} \left\{ |k \rangle \otimes
\begin{pmatrix}
1 \\
0
\end{pmatrix}  \pm
|k-1 \rangle \otimes
\begin{pmatrix}
0 \\
1
\end{pmatrix}  \right\},
\end{equation}
with energy $E^a_k = \hbar \omega k$ whereas for $\mathcal{H}^b = \mathcal{F}^b \otimes \mathbb{C}^2$ and the Hamiltonian $H^b  = \hbar \omega \{ Q^b, Q^{b \dagger} \}$, one has
\begin{equation}
\left\{ |l-1 \rangle \otimes
\begin{pmatrix}
1 \\
0
\end{pmatrix} ,
|l \rangle \otimes
\begin{pmatrix}
0  \\
1
\end{pmatrix}
\right\},
\end{equation}
and the (maximally) entangled state
\begin{equation}
\dfrac{1}{\sqrt{2}}
\begin{pmatrix}
|l-1 \rangle \\
\pm |l \rangle
\end{pmatrix} =
\dfrac{1}{\sqrt{2}} \left\{ |l-1 \rangle \otimes
\begin{pmatrix}
1 \\
0
\end{pmatrix}  \pm
|l \rangle \otimes
\begin{pmatrix}
0 \\
1
\end{pmatrix}  \right\},
\end{equation}
with energy $E^b_l = \hbar \omega l$. 

Denoting the composite bosonic part as $\mathcal{F}=\mathcal{F}^a \otimes \mathcal{F}^b$ analogous to (15), we wish to analyze the following entangled supermultiplet state in $ \mathcal{F} \otimes \mathbb{C}^2$, 
\begin{equation}
| \Phi_k \rangle = \alpha \, | k \,\, k-1 \rangle \otimes \begin{pmatrix}
1 \\
0
\end{pmatrix} + \beta \, | k-1 \,\, k \rangle \otimes \begin{pmatrix}
0 \\
1
\end{pmatrix} .
\end{equation}
The entanglement of formation is directly related to the concurrence measure as stated in the appendix. The concurrence of $| \Phi_k \rangle $, $C(| \Phi_k \rangle )$, is given by the absolute value of the expectation value of the modular conjugation operator in this state, 
\begin{equation}
C(| \Phi_k \rangle ) = | \langle \Phi_k | J | \Phi_k \rangle |.
\end{equation}
Using (34), we have
\begin{equation}
 J | \Phi_k \ \rangle =  \bar{\alpha} | k-1 \,\,k \rangle \otimes \begin{pmatrix}
0 \\
1
\end{pmatrix} + \bar{\beta} |k \,\, k-1 \rangle \otimes \begin{pmatrix}
1 \\
0
\end{pmatrix} ,
\end{equation} 
and therefore, 
\begin{equation}
\langle \Phi_k | J | \Phi_k \ \rangle = 2\bar{\alpha}\bar{\beta} , \;\;\;\;
| \langle \Phi_k | J | \Phi_k \rangle |= 2 |\alpha \beta |,
\end{equation}
which is the concurrence using traditional methods \cite{Ben2019}. For  bipartite states, the entanglement of formation is directly related to the concurrence of the state, $C(| \Phi \rangle)$. For a maximally entangled state with $\alpha = \beta = 1/\sqrt{2}$, $C(| \Phi_k \rangle ) = 1$ as required. The concurrence relation (40) has given a physical meaning to the modular conjugation operator as a quantitative measure of entanglement for bi-partite supermultiplet states. As described in \cite{Uhlmann2000}, an anti-linear, anti-unitary operator is the key driving force behind the entanglement of formation of bipartite systems and it is now understood in this context as the modular conjugation operator $J$.

\section{Applications to 2D Dirac Fermions in Graphene} 
Massless Dirac fermions restricted to two spatial dimensions appear as low energy electrons in graphene, a layer of one-atom thick carbon atoms in a honeycomb lattice. A low energy Hamiltonian valid around two Dirac points $\xi = \pm $ in the first Brillouin zone with a perpendicular magentic field is given by \cite{Ezawa2013,Dup2017}
\begin{equation}
H_D^{\xi} = v_{F} (\xi \hat{\sigma}_x \Pi_x + \hat{\sigma}_y \Pi_y) ,
\end{equation}
where $v_F $ is the Fermi velocity. Using (2) and (6), one has
\begin{equation}
H_D^{a +} = \hbar \omega_D (Q^a + Q^{a \dagger}) = \hbar \omega_D Q^a_{SUSY}=  \hbar \omega_D 
\begin{pmatrix}
0 & a^\dagger \\
a & 0
\end{pmatrix} ,
\end{equation}
and
\begin{equation}
H_D^{a -} = \hbar \omega_D (Q^{a T} + Q^{a \dagger T}) = \hbar \omega_D Q^{a T}_{SUSY} = \hbar \omega_D 
\begin{pmatrix}
0 & a\\
a^\dagger & 0
\end{pmatrix},
\end{equation}
where $\omega_D = \sqrt{2} \hbar v_F/l_m$. Both these Hamiltonians are scaled supercharges. This system will have the same degeneracy as the system in the previous section as these Hamiltonians commute with the '$b$' operators (13). Therefore, the respective eigenvectors and eigenvalues are, using (8), (9), and (15) 
\begin{equation}
\begin{pmatrix}
|n, m \rangle \\
|n-1, m \rangle
\end{pmatrix} , \,\,\,
\begin{pmatrix}
|n-1, m \rangle \\
|n, m \rangle
\end{pmatrix}, \;\;\;
E_n = \hbar \omega_D \sqrt{n} = \hbar^2 \dfrac{v_F}{l_m} \sqrt{2n}, n \in \mathcal{Z}^+ .
\end{equation}
Using the modular conjugation operator  $J \mathcal{A} J = \mathcal{A}'$, one can explicitly execute the symmetry into the "$b$" space where the $\mathbf{B}$ field points in the reverse direction, i.e. $J H_D^{a +} J = H_D^{b -}$, resulting in Hamiltonians
\begin{equation}  
\begin{array}{c}
H_D^{b +} = \hbar \omega_D
\begin{pmatrix}
0 & b^\dagger \\
b & 0
\end{pmatrix} 
,
H_D^{b -} = \hbar \omega_D
\begin{pmatrix}
0 & b \\
b^\dagger & 0
\end{pmatrix} ,\end{array}
\end{equation}
and their respective eigenvectors and eigenvalues
\begin{equation}  
\begin{array}{c}
\begin{pmatrix}
|n, m \rangle \\
|n, m-1 \rangle
\end{pmatrix} , \,\,\,
\begin{pmatrix}
|n, m-1 \rangle \\
|n, m \rangle
\end{pmatrix} , \;\;\;
E_m = \hbar \omega_D \sqrt{m} = \hbar^2 \dfrac{v_F}{l} \sqrt{2m}, m \in \mathcal{Z}^+.
\end{array}
\end{equation}

Note that using the supersymmetric algebra of (6) and (7), one may consider a map to the structure of the previous section,
\begin{equation}
\begin{array}{c}
(H_D^+)^2/(\hbar \omega_D)^2 = \{Q^a,Q^{a \dagger} \} =  (Q^{a}_{SUSY})^2= H^a/{\hbar \omega}, \\
(H_D^-)^2/(\hbar \omega_D)^2 = \{Q^{a T},Q^{a \dagger T} \} =  (Q^{a T}_{SUSY})^2 .
\end{array}
\end{equation}
This has the structure of a nonlinear SUSY  algebra (see \cite{And2003, And2012}).

\section{Modular Operators and Concurrence in the Jaynes Cummings Model} 

A further example of a SUSY quantum mechanical system is the Jaynes–Cummings model of quantum optics. This model describes the system of a two-level atom interacting with the quantized mode of an optical cavity (photons). The Hamiltonian in resonance and within the rotating wave approximation (RWA) is given by \cite{Lou1973} 
\begin{equation}
H_{JC}^b = \hbar \omega  \left( b^\dagger b \otimes I + I \otimes \dfrac{\sigma_z}{2} \right) + \hbar g (b \otimes \sigma_+ + b^\dagger \otimes \sigma_{-} ),
\end{equation} 
or upon using (16) and (17) it's SUSY form,
\begin{equation}
H_{JC}^b = \hbar \omega \left\{ Q^b , Q^{b \dagger} \right\} + \hbar g (Q^b + Q^{b \dagger} ) -\dfrac{\hbar \omega }{2}.
\end{equation} 
The conserved supercharge operator that creates an energy degeneracy via supersymmetry, $Q^b_{SUSY}=Q^b+ Q^{b\dagger}$ allows one to write the Hamiltonian as
\begin{equation}
H_{JC}^b = \hbar \omega \{Q^b_{SUSY}\}^2+ \hbar g Q^b_{SUSY} -\dfrac{\hbar \omega }{2}.
\end{equation} 
The (maximally) entangled state
\begin{equation}
\dfrac{1}{\sqrt{2}}
\begin{pmatrix}
|l-1 \rangle \\
\pm |l \rangle
\end{pmatrix} =
\dfrac{1}{\sqrt{2}} \left\{ |l-1 \rangle \otimes
\begin{pmatrix}
1 \\
0
\end{pmatrix}  \pm
|l \rangle \otimes
\begin{pmatrix}
0 \\
1
\end{pmatrix}  \right\}
\end{equation}
has a SUSY energy given by $E^{b}_{SUSY} =  \hbar \omega l  \pm \hbar g \sqrt{l} -\dfrac{\hbar \omega}{2}$.

Applying the results  of section III, the modular conjugation operator $J$ maps (50) to a new system with a Hamiltonian given by  
\begin{equation}
H_{AJC}^a = \hbar \omega  \left( a^\dagger a \otimes I - I \otimes \dfrac{\sigma_z}{2} \right) + \hbar g (a \otimes \sigma_{-} + a^\dagger \otimes \sigma_{+} ),
\end{equation} 
which is an anti-Jaynes Cummings (AJC) model (\cite{Dod2019,Chor2017}) with only the non-RWA interaction term. Upon using (6) and (7) it's SUSY form is
\begin{equation}
H_{AJC}^a = \hbar \omega \left\{ Q^a , Q^{a \dagger} \right\} + \hbar g (Q^a + Q^{a \dagger} ) -\dfrac{\hbar \omega }{2}.
\end{equation} 
The conserved supercharge operator that creates an energy degeneracy via supersymmetry, $Q^a_{SUSY}=Q^a+ Q^{a\dagger}$ allows one to write the Hamiltonian as
\begin{equation}
H_{AJC}^a = \hbar \omega \{Q^a_{SUSY}\}^2+ \hbar g Q^a_{SUSY} -\dfrac{\hbar \omega }{2}.
\end{equation} 
The (maximally) entangled state
\begin{equation}
\dfrac{1}{\sqrt{2}}
\begin{pmatrix}
|k \rangle \\
\pm |k-1 \rangle
\end{pmatrix} =
\dfrac{1}{\sqrt{2}} \left\{ |k \rangle \otimes
\begin{pmatrix}
1 \\
0
\end{pmatrix}  \pm
|k-1 \rangle \otimes
\begin{pmatrix}
0 \\
1
\end{pmatrix}  \right\}
\end{equation}
has a SUSY energy given by $E^{a}_{SUSY} = \hbar \omega k  \pm \hbar g \sqrt{k} -\dfrac{\hbar \omega}{2}$.

\section{Conclusion}
The Tomita Takesaki modular operator approach to von Neumann algebras has been elucidated in the algebraic structure of certain SUSY quantum mechanical systems. Explicit modular operators have been given for these systems along with their physical descriptions.
The concurrence measure used to calculate the entanglement of formation has been known to be related to an anti-linear, anti-unitary operator. Using SUSY quantum mechanics, we have shown that this operator is the modular conjugation operator of the Tomita Takesaki modular operator formalism of the von Neumann algebraic structure of the SUSY system. This concurrence measure has been demonstrated on entangled supermultiplet states. The use of the  modular operator formalism has provided both a deeper mathematical meaning to concurrence while proving a direct physical meaning to the modular conjugation symmetry operator of the SUSY system. Finally, the theory was applied to the case of two-dimensional Dirac fermions, as is found in graphene, and a  supersymmetric Jaynes Cummings Model.

Our hope is that the results presented in this paper may lead to further investigations and extensions in the connection between von Neumann algebras and SUSY quantum mechanics. For instance, supersymmetric theories with topological defects can have nontrivial behaviors determined by whether or not supersymmetry is restored in the defect core \cite{Koehn2016, Oikonomou2014II}. It would be interesting to see how this symmetry breaking affects any possible von Neumann algebra symmetries present in these systems. Furthermore, our study may provide elucidation to infinite systems such as those of algebraic quantum field theory  whereby the relative modular operator of a von Neumann algebra of quantized fields has been used as a measure of relative entanglement entropy. Further details along this direction may be found in \cite{Ho2018, Witt2018}.   

\section{Appendix}
For more details, see \cite{KR1983,Tak1979,Brat1987,Junk1996,Ben2019, Woot1998}.

\textit{\textbf{Banach Algebra:}} Let $\mathcal{A}$ be an algebra. A normed algebra has a norm map: $\mathcal{A} \rightarrow \mathbb{R}^+ , \,\,\, A \rightarrow ||A|| \in  \mathbb{R}^+ , \forall A \in \mathcal{A} $ such that $||A|| \geq 0,||A|| = 0 \iff A=0, \alpha \in \mathbb{C} , ||\alpha A || = |\alpha| \, ||A||,$ $|| A+B|| \leq ||A|| + ||B||, \; and \;||AB|| \leq ||A|| \, ||B|| .$ A Banach algebra is a complete normed algebra (complete in the norm map).

\textit{\textbf{$C^*$-Algebra:}} A $C^*$-algebra $\mathcal{A}$ is a Banach algebra with an involutive map $ ^*: \mathcal{A} \rightarrow \mathcal{A} , \,\,\, A \rightarrow A^* \,\,\, \forall A \in \mathcal{A} , \lambda \in \mathbb{C}$ such that

$(A^*)^*=A,\;\;\; (AB)^*=B^* A^*,\;\;\;(\lambda A)^*= \bar{\lambda} A^*,\;\;\; (\lambda  A + \alpha B )^* = \bar{\lambda } A^* +\bar{\alpha}  B^* , $

$||A^*|| = ||A||,\; \;\;\; ||A A^*|| = ||A|| \, ||A^*||$ ($C^*$ condition).

\textit{\textbf{Von Neumann Algebra:}} Consider a  $C^*$-algebra $\mathcal{B(H)} =\{A\}$ of bounded linear operators on a Hilbert space, $ A: \mathcal{H} \rightarrow \mathcal{H}$. Let $\mathcal{C}$ be a subset of $\mathcal{B(H)}$. An operator $A \in \mathcal{B(H)} $ belongs to the commutant $\mathcal{C}'$ of the set $\mathcal{C}$ $\iff$   $AC =CA, \,\,\, \forall C \in \mathcal{C}$. A von Neumann algebra $\mathcal{A}$ is a unital $C^*$-subalgebra of $\mathcal{B(H)}$ such that $\mathcal{A}'' = \mathcal{A}$. Consider a von Neumann algebra $\mathcal{A} \subset \mathcal{B(H)}$. A von Neumann algebra in standard form is one where there exists an element $| \Omega \rangle \in \mathcal{H}$ which is both cyclic (operating on $| \Omega \rangle $ with elements in  $\mathcal{A}$ can generate a space dense in $\mathcal{H}$) and separating (if $A | \Omega \rangle = 0$, then $A=0$).

\textit{\textbf{Tomita Takesaki Modular Operators:}} Consider a  von Neumann algebra $\mathcal{A} \subset \mathcal{B(H)}$ in standard form with a cyclic and separating vector $| \Omega \rangle \in \mathcal{H}$. Let $S:\mathcal{H} \rightarrow \mathcal{H}$ be a anti-unitary operator defined by $S A  | \Omega \rangle = A^* | \Omega \rangle $. Let the closure of $S$ have a polar decomposition given by $S=J \Delta ^{\frac{1}{2}} =\Delta^{-\frac{1}{2}} J$, where $J$ is called the modular conjugation operator and $\Delta$ is called the modular operator. $J$ is anti-linear and anti-unitary whereas $\Delta$ is self-adjoint and positive. Furthermore, the following relations hold:

1. $J \Delta ^{\frac{1}{2}} J = \Delta ^{-\frac{1}{2}}\;\;\;\;$
2. $ J^2 =I, \,\,\, J^{*} = J \;\;\;\;$
3. $ J |\Omega \rangle  = |\Omega \rangle\;\;\;\;$
4. $ J \mathcal{A} J = \mathcal{A}' $

5. $ \Delta = S^{*} S\;\;\;\;$
6. $ \Delta |\Omega \rangle  = |\Omega \rangle$

7. $ \Delta^{it} \mathcal{A} \Delta^{-it} = \mathcal{A} \;\;\;\;$ (one parameter-$t$ group of automorphisms of $\mathcal{A})$

8. If $ \omega (A) = \langle \Omega | A \Omega \rangle $, $ \forall A \in \mathcal{A}$,  then $\omega$ is a KMS (Kubo-Martin-Schwinger) functional (state) on $\mathcal{A}$ with respect to the automorphism of 7.

9.$|\Omega \rangle $  is cyclic for $\mathcal{A}$ if and only if $|\Omega \rangle $  is separating for $\mathcal{A}'$

\textit{\textbf{SUSY Quantum Mechanics:}} A supersymmetric quantum mechanical system is characterized by a set of operators $\{H, Q_1, Q_2, ..., Q_N \}$ with the following anti-commutation relations:

1. $\{Q_i,Q_j^{\dagger}\}=H \delta_{ij}, \,\,\, i,j =1,2,...N\;\;\;\;$ 
2. $\{Q_i,Q_i\}=\{Q_i^{\dagger},Q_i^{\dagger}\}=0, \,\,\, i =1,2,...N$

Using 2., one can easily show that the supercharges $Q_i$ are conserved quantities, i.e.
\begin{equation*}
[H, Q_i] =  [Q_iQ_i^{\dagger}, Q_i] + [Q_i^{\dagger}Q_i, Q_i] 
=Q_i[Q_i^{\dagger}, Q_i] + [Q_i^{\dagger}, Q_i]Q_i 
=Q_iQ_i^{\dagger}Q_i - Q_iQ_i^{\dagger}Q_i =0  .
\end{equation*}

\textit{\textbf{ Entanglement of Formation (Wooters \cite{Woot1998}) and Concurrence:}} The entanglement of formation of a generic bi-partite state $\rho$ (quantum density operator) is related to the concurrence of that state $C(\rho)$ as follows:
\begin{equation*}
E_{Formation}(\rho)=H_{bin}\left( \dfrac{1+\sqrt{1+C(\rho)}}{2} \right)
\end{equation*}
where $H_{bin}$ is the Shannon binary entropy.

\vspace{10mm}
Some basic properties for $J$ stated in (23) are verified as follows. 

\textbf{(i)} Clearly, $J^2 = I$. We need to show the $J^*=J^{\dagger}=J$. 
 
Let $| \psi \rangle = |n , m \rangle \otimes \left\{ \alpha  
\begin{pmatrix}
1 \\
0
\end{pmatrix} +\beta 
 \begin{pmatrix}
0 \\
1
\end{pmatrix} \right\}$ such that $\langle \psi | \psi \rangle  =1$. 
A straightforward calculation shows that
\begin{equation}
\langle \psi | J^\dagger J | \psi \rangle = \langle m, n| m, n \rangle \left( |\alpha|^2 + |\beta|^2 \right) = 1 = \langle \psi | \psi \rangle .
\end{equation}
Since $J^2 =1$, $J^\dagger  = J$. 

\textbf{(ii)} $J$ is \textit{anti-unitary}. 

Let
\begin{equation}
| \xi \rangle = |k ,l \rangle \otimes \left\{ \delta  
\begin{pmatrix}
1 \\
0
\end{pmatrix} +\gamma 
 \begin{pmatrix}
0 \\
1
\end{pmatrix} \right\} .
\end{equation}
Since,
\begin{equation}
J | \psi \rangle = | m ,n \rangle \otimes \left\{ \bar{\alpha}  
\begin{pmatrix}
0 \\
1
\end{pmatrix} +\bar{\beta} 
 \begin{pmatrix}
1 \\
0
\end{pmatrix} \right\} ,
\;\;\;
J | \xi \rangle = |l , k \rangle \otimes \left\{ \bar{\delta}  
\begin{pmatrix}
0 \\
1
\end{pmatrix} +\bar{\gamma} 
 \begin{pmatrix}
1 \\
0
\end{pmatrix} \right\} ,
\end{equation}
anti-unitarity follows as $\langle J \xi | J \psi \rangle = \langle l | m \rangle \langle k | n \rangle [ \delta \bar{\alpha} + \gamma \bar{\beta}] =\delta_{ l  m } \delta_{k  n}  [ \delta \bar{\alpha} + \gamma \bar{\beta} ] = \langle  \psi |  \xi \rangle $.
 
\textbf{(iii)} $J$ maps $\mathcal{A}$ into $\mathcal{A}'$, i.e. $J \mathcal{A} J = \mathcal{A}'$.
Using (34), we have 
\begin{equation}
\begin{array}{c}
J Q^a J 
| \psi \rangle
=
J \begin{pmatrix}
0 & 0\\
a & 0
\end{pmatrix}
 |m , n \rangle \left\{ \bar{\alpha} 
\begin{pmatrix}
0 \\
1
\end{pmatrix} +\bar{\beta} 
 \begin{pmatrix}
1 \\
0
\end{pmatrix} \right\} = 
J \bar{\beta} 
\begin{pmatrix}
0 \\
a |m , n \rangle
\end{pmatrix} \\
\\
= 
J \bar{\beta} 
\begin{pmatrix}
0 \\
\sqrt{m} |m-1 , n \rangle
\end{pmatrix} 

=
\beta 
\begin{pmatrix}
\sqrt{m}  |n , m-1 \rangle \\
0
\end{pmatrix} =
\beta 
\begin{pmatrix}
b  |n , m \rangle \\
0
\end{pmatrix} \\ 
\\
=
\begin{pmatrix}
0 & b\\
0 & 0
\end{pmatrix}
 |n , m \rangle \left\{ \alpha
\begin{pmatrix}
1 \\
0
\end{pmatrix} +\beta 
 \begin{pmatrix}
0 \\
1
\end{pmatrix} \right\}
 = 
Q^b |n , m \rangle \otimes \left\{ \alpha
\begin{pmatrix}
1 \\
0
\end{pmatrix} +\beta 
 \begin{pmatrix}
0 \\
1
\end{pmatrix} \right\}  = Q^b |\psi \rangle .
\end{array}
\end{equation}
A similar calculation holds for the other operators.

\textbf{(iv)} \textit{Invariance of} $|\Omega \rangle$ 

The normalized separating and cyclic vector  $ |\Omega \rangle $ is explicitly given by ($\beta \in \mathbb{R}$)
\begin{equation}
|\Omega \rangle = [1-e^{-\beta}]^{\frac{1}{2}} \sum_{n=0}^{\infty} e^{-\frac{\beta n}{2}} | n, n \rangle \otimes  \Bigg\{
\begin{pmatrix}
1\\
0
\end{pmatrix}  +  
\begin{pmatrix}
0\\
1
\end{pmatrix}  \Bigg\}
\end{equation} 
By using property 9 of the Tomita-Takesaki modular operators, $|\Omega \rangle $ is  separating in $\mathcal{A}'$ and therefore is cyclic in $\mathcal{A}$ and vice versa (\cite{Ali2010}). It is clearly invariant under $J$, $J |\Omega \rangle= |\Omega \rangle$. 

\textbf{(v)} \textit{The modular operator} $\Delta$ and $J \Delta ^{\frac{1}{2}} J = \Delta ^{-\frac{1}{2}} $. 
The modular operator is given by  
\begin{equation}
\Delta = \exp[-\beta (H^a - H^b)] 
\end{equation}
Using (iii) above, one has
\begin{equation}
J \Delta^{1/2}J = J \exp[-\beta (H^a - H^b)/2]J 
=\exp[-\beta (H^b - H^a)/2]=\Delta^{-1/2}.
\end{equation}

\textbf{(vi)} \textit{The anti-linear operator} $ S A |\Omega \rangle  = A^* |\Omega \rangle , \,\,\, S = J \Delta^{1/2}$ (set $\hbar \omega =1$)
\begin{equation}
\begin{array}{c}
S Q^a |\Omega \rangle = J \Delta^{1/2} Q^a |\Omega \rangle 
=J \Delta^{1/2}  \begin{pmatrix}
0 & 0\\
a & 0
\end{pmatrix}  [1-e^{-\beta}]^{\frac{1}{2}} \displaystyle\sum_{n=0}^{\infty} e^{-\frac{\beta n}{2}} | n,n \rangle \otimes  \Bigg\{
\begin{pmatrix}
1\\
0
\end{pmatrix}  +  
\begin{pmatrix}
0\\
1
\end{pmatrix}  \Bigg\}  \\
= J \Delta^{1/2} [1-e^{-\beta}]^{\frac{1}{2}}  \displaystyle\sum_{n=0}^{\infty} e^{-\frac{\beta n}{2}} 
\begin{pmatrix}
0\\
a | n,n \rangle 
\end{pmatrix} 
 = J\Delta^{1/2} [1-e^{-\beta}]^{\frac{1}{2}}  \displaystyle\sum_{n=1}^{\infty} e^{-\frac{\beta n}{2}} 
\begin{pmatrix}
0\\
\sqrt{n} | n-1, \, n \rangle 
\end{pmatrix}    \\ 
=  J  [1-e^{-\beta}]^{\frac{1}{2}}  \displaystyle\sum_{n=1}^{\infty} e^{-\frac{\beta n}{2}} \exp\left[-\dfrac{\beta}{2}
\begin{pmatrix}
a^{\dagger}a-bb^{\dagger} & 0 \\
0 & a a^{\dagger}-b^{\dagger}b
\end{pmatrix}
\right]
\begin{pmatrix}
0\\
\sqrt{n} | n-1, \, n \rangle 
\end{pmatrix} \\ 
=  J [1-e^{-\beta}]^{\frac{1}{2}}  \displaystyle\sum_{n=1}^{\infty} e^{-\frac{\beta n}{2}} \exp[-\beta (n-n)/2]
\begin{pmatrix}
0\\
\sqrt{n} | n-1, \, n \rangle 
\end{pmatrix} \\
=  J  [1-e^{-\beta}]^{\frac{1}{2}}  \displaystyle\sum_{n=1}^{\infty} e^{-\frac{\beta n}{2}}
\sqrt{n} | n-1 ,\,  n \rangle  \otimes 
\begin{pmatrix}
0\\
1 
\end{pmatrix} 
= [1-e^{-\beta}]^{\frac{1}{2}}  \displaystyle\sum_{n=1}^{\infty} e^{-\frac{\beta n}{2}}
\sqrt{n} | n ,\,  n-1 \rangle  \otimes 
\begin{pmatrix}
1\\
0 
\end{pmatrix} \\
=\begin{pmatrix}
0 & b\\
0 & 0
\end{pmatrix}  [1-e^{-\beta}]^{\frac{1}{2}} \displaystyle\sum_{n=0}^{\infty} e^{-\frac{\beta n}{2}} | n,n \rangle \otimes  \Bigg\{
\begin{pmatrix}
1\\
0
\end{pmatrix}  +  
\begin{pmatrix}
0\\
1
\end{pmatrix}  \Bigg\} = Q^b |\Omega \rangle .
\end{array}
\end{equation}
The other generators follow in a similar manner.

\end{document}